\documentclass{PoS}
\usepackage{amssymb,amsmath}

\title{Multigrid Algorithms for Domain-Wall Fermions}
\ShortTitle{Multigrid Algorithms for Domain-Wall Fermions}

\author{\speaker{Saul D. Cohen} \\
        Department of Physics, University of Washington, Seattle, WA 98195 \\
        E-mail: \email{sdcohen@uw.edu}}
        
\author{R.~C.~Brower \\
        Center for Computational Science, Boston University, Boston, MA 02138}

\author{M.~A.~Clark \\
        Harvard-Smithsonian Center for Astrophysics, Harvard University, Boston, MA 02215}

\author{J.~C.~Osborn \\
        Argonne Leadership Computing Facility, Argonne National Laboratory, Argonne, IL 60439, USA}

\abstract{We describe an adaptive multigrid algorithm for solving inverses of the domain-wall fermion operator. Our multigrid algorithm uses an adaptive projection of near-null vectors of the domain-wall operator onto coarser four-dimensional lattices. This extension of multigrid techniques to a chiral fermion action will greatly reduce overall computation cost, and the elimination of the fifth dimension in the coarse space reduces the relative cost of using chiral fermions compared to discarding this symmetry. We demonstrate near-elimination of critical slowing as the quark mass is reduced and small volume dependence, which may be suppressed by taking advantage of the recursive nature of the algorithm.}

\FullConference{The XXIX International Symposium on Lattice Field Theory\\
		 2011 July 10--16\\
		 Lake Tahoe, CA, USA}
		
\begin{document}

\section{Introduction}
Lattice quantum chromodynamics (QCD) is a method for deriving the consequences of QCD in nonperturbative regimes, where standard analytic techniques cannot be applied. Since lattice QCD directly uses the equations of QCD without recourse to modeling, it offers straightforward controlled analysis of the uncertainties of its results. One significant outstanding uncertainty in current lattice calculations is the use of unphysically heavy masses for the up and down quarks, which must be extrapolated out at the end of a calculation. The use of heavy quark masses is motivated by the divergent increase in computational cost as the quark mass approaches zero. Although the up and down quarks have small nonzero masses, the computational power needed to use lattice QCD at such small masses has not been available. Even today, as the explosion of available computational power has made running at the physical quark masses seem practical for the first time, methods for reducing the amount of computation are highly desirable.

The root cause of the critical slowing is the structure of the Dirac operator.
Since the smallest eigenvalue of the Dirac operator is linear in the quark mass, as the quark mass approaches zero, so do the lowest modes of the operator.
This means that the condition number of the operator will diverge, greatly slowing the convergence of the Krylov-subspace solvers used to find the quark propagator. An obvious method for attacking this problem is to find a simple transformation of the operator that decreases its condition number; such techniques are called preconditioning. A careful balance must be struck between the quality of the preconditioner, which saves computational cost in the top-level solver, and the complexity of the preconditioner, which has its own intrinsic computational cost. In addition, certain preconditioning techniques will perform better for certain problems or have costs that scale differently with the volume of the lattice.

Many preconditioning techniques have met with some success in lattice QCD, such as even-odd preconditioning. One technique of significant interest is deflation\cite{Morgan:2007dq,Luscher:2007se,Stathopoulos:2007zi}. Eigenvector deflation takes advantage of the fact that the Dirac operator has a small number of low modes relative to the size of the matrix. By explicitly computing these low modes, they can be projected out of the operator, improving its condition number. However, the number of low modes on the lattice is expected to scale roughly as its volume; this means that the amount of computational work needed to use deflation as a preconditioner will scale linearly with the volume. Since physical quark masses will yield a very light pion, a lattice volume must be quite large to manage finite-volume effects, which scale as $e^{-M_\pi L}$. Thus, deflation will become costly even as we enter the part of the parameter space where preconditioning is most needed.

In many other fields of applied mathematics, the preconditioner of choice for solving discretized differential equations is the multigrid. The key idea behind the multigrid may be understood by considering the action of a Krylov solver on the residual $r = Ax-b$, where $b$ is some source vector, $A$ is the operator of interest and $x$ is a proposed solution. During each step of the Krylov solver, the solution vector $x$ is updated by some amount that zeros the residual or conjugate residual along some search direction in the Krylov space. The Krylov space is populated by vectors multiplied by powers of the matrix, so high eigenmode components of the vectors are enhanced. As a result, the residual will quickly be depleted of high (short-wavelength) modes, but the lower modes will be little affected by the updates; the Krylov solver suffers from critical slowing. However, the scale at which we have discretized the problem is to some degree arbitrary. We are free to invent a new coarser grid on which the long-wavelength modes are no longer so long. Operating on this coarser grid, the same Krylov solver will be far more effective in eliminating the low modes from the residual. If we find that the coarse-grid solver is still unacceptably slow, the entire procedure may be repeated on yet coarser grids, until the critical slowing is eliminated.

Defining exactly what we mean by ``inventing a coarser grid'' means introducing just two new operators: a restrictor $R$ which takes fine-grid vectors and returns coarse-grid vectors and a prolongator $P$ which works vice-versa. For many problems in applied mathematics, such as the Laplacian operator, an excellent restrictor is to simply divide the fine lattice into non-overlapping subsets and average the fine vectors over each subset to make one site of the coarse vector. We define the prolongator by the condition $RP = 1_c$, where $1_c$ denotes the identity on the coarse grid. Such techniques are known as geometric multigrid algorithms.

Although quite helpful on problems where the low modes are smooth long-wavelength objects, algebraic multigrid did not help much in lattice QCD. The reason for this is that the low modes of QCD do not match the criteria of smooth and long-wavelength. Rather, they are blobby objects with characteristic scales of the size of instantons. Since algebraic multigrid fails due to our preconception of the shapes of the low modes, a technique that accounts for their true shape would be a natural extension. Instead of trying to specify such shapes ourselves, we can identify problematic low modes using the definition above: those modes that are not removed quickly by our Krylov solver. Techniques falling into this category are known as adaptive multigrid algorithms.

In our previous work\cite{Osborn:2010mb,Babich:2010qb}, we demonstrated the efficacy of adaptive multigrid algorithms in eliminating critical slowing for the Wilson Dirac operator. In this proceeding, we will show how adaptive multigrid may be applied to domain-wall fermions. In Section~\ref{sec:alg}, we will give a detailed description of the algorithm, explicitly indicating how to form the prolongator and restrictor, and indicating which Krylov solvers perform well. In Section~\ref{sec:num}, we apply the algorithm to the toy problem of U(1) gauge fields in 2 dimensions as well as to actual lattice ensembles currently in use. We demonstrate that the algorithm eliminates the critical slowing at low quark mass. We discuss briefly the cost-benefit of using the multigrid algorithm.

\section{Description of the Algorithm}\label{sec:alg}

The essence of the multigrid algorithm is the reduction of a problem to a coarser grid where it can be solve more effectively. We can visualize this process as a ``descent'' to a coarser grid, followed by a reascent to the fine grid, a process known as a V-cycle. A V-cycle applied to solving the problem $Ax=b$ consists of five steps:
\begin{itemize}
\item Apply $\nu_1$ iterations of a solver $S_1$ at the fine level: $x_1 = S_1^{\nu_1}(A,b)$.
\item Restrict the residual to the coarse lattice: $b_2 = R(b-Ax_1)$.
\item Apply $\nu_2$ iterations of a solver $S_2$ at the coarse level: $x_2 = S_2^{\nu_2}(RAP,b_2)$.
\item Prolong the correction, accumulate with our previous solution and determine the corrected residual: $b_3 = b - A(Px_2 + x_1)$.
\item Apply $\nu_3$ iterations of a solver $S_3$ at the fine level: $x = x_1 + Px_2 + S_3^{\nu_3}(A,b_3)$.
\end{itemize}
Expressed in this form, we can immediately see that the algorithm is highly modular and lends itself to recursion. In particular, the solver $S_2$ may itself be a V-cycle, connecting the coarse problem to even coarser grids. In other cases, we may replace $S_2$ by two V-cycles, forming a three-level process called a W-cycle. Further generalizations are also possible.

One important result of this recursive property is that the multigrid algorithm can remove critical slowing in all cases. In such cases where a one-level multigrid solver is unable to remove critical slowing, due to divergence of the number of iterations needed at the coarse level, another coarser level can be added.

Finding the best subspace for the coarse level is an important part of the multigrid algorithm. The most obvious subspace to use is one containing the eigenvectors corresponding to the smallest eigenvalues of the operator. These vectors are simply divided up into blocks corresponding to the coarse-lattice sites, and the coarse-level subspace consists of the entire subspace spanned by these blocks of the vectors. This is a much larger space than the one spanned by the vectors themselves, and from the algebraic multigrid derives some of its power from this expansion. The coarse-level subspace will, in addition to the lowest eigenvectors, also contain the greater part of many of the low eigenvectors not explicitly included.

\begin{figure}
\begin{center}
\includegraphics[width=0.4\textwidth]{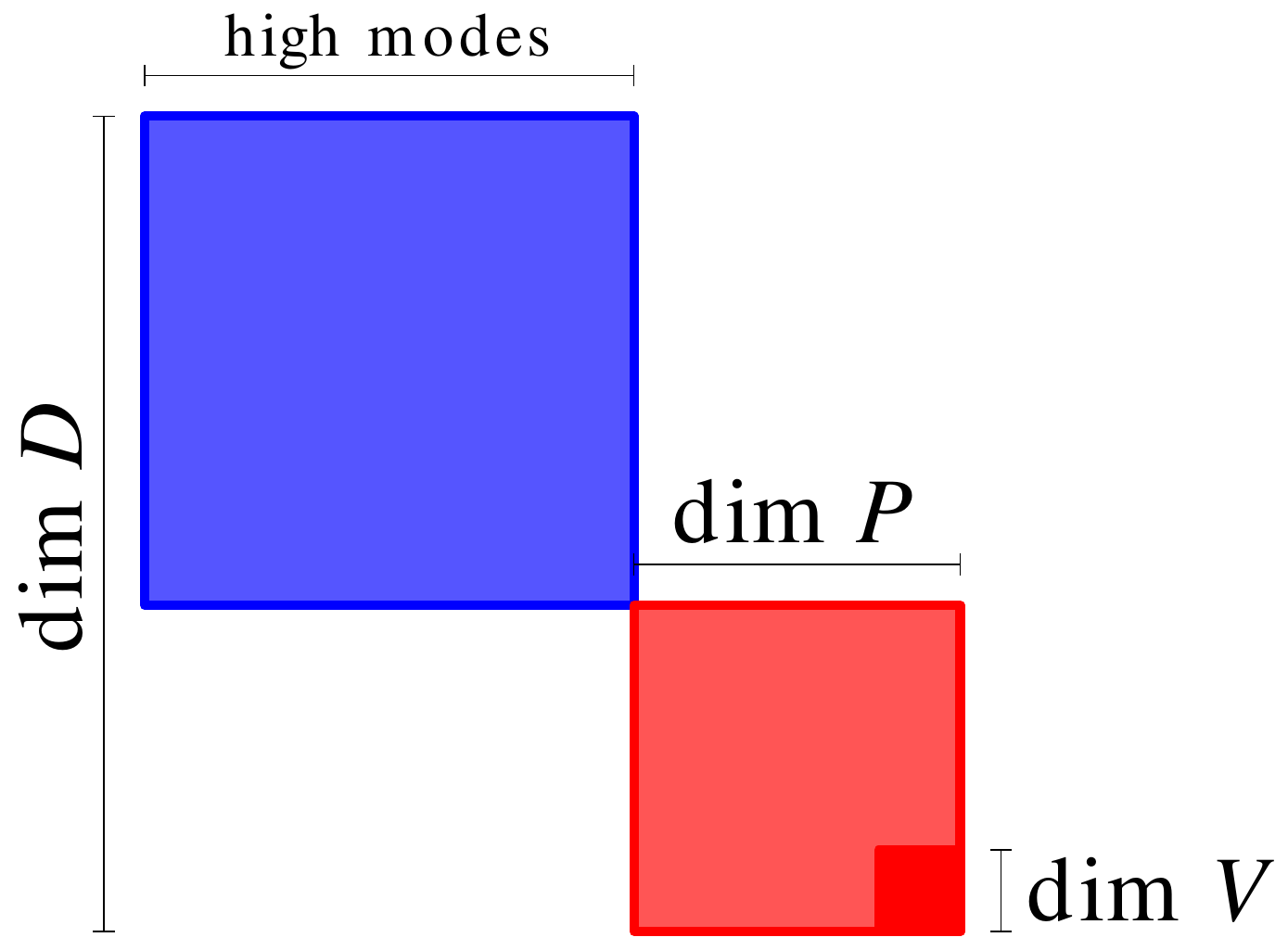}
\includegraphics[width=0.55\textwidth]{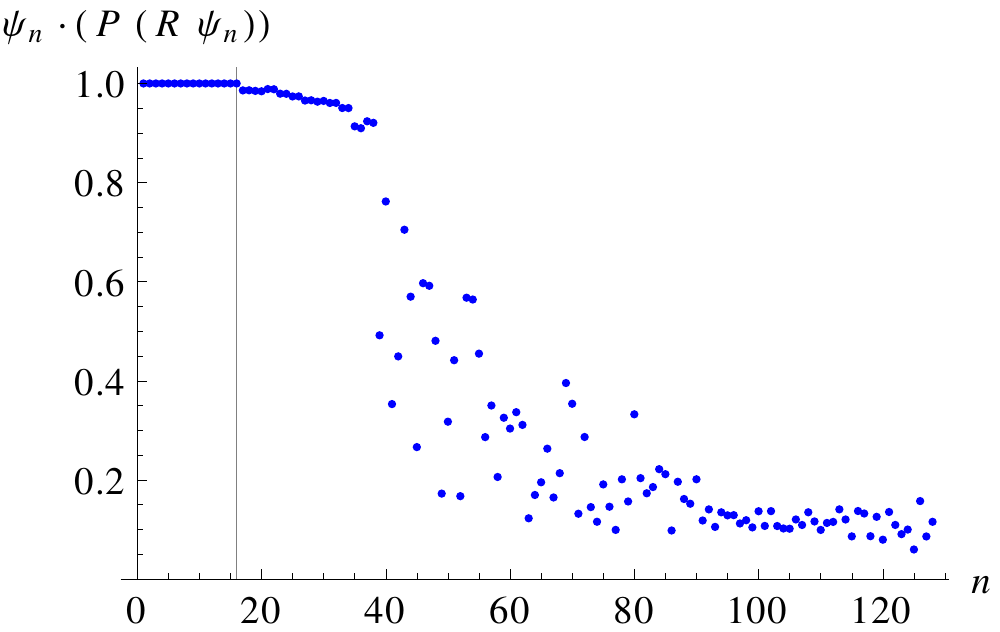}
\end{center}
\caption{\label{fig:nullspace} (Left) A conceptual depiction of how multigrid divides the vector space. The dark red space contains eigenmodes included in the explicitly derived null space; such modes would be removed by deflation-type schemes. The light red space contains additional low modes included in the multigrid's coarse-lattice space. The blue space contains high modes not included at the coarse level but well converged by the fine-space Krylov solver. (Right) The overlap of the low eigenmodes of the Dirac operator with the coarse space. The 16 modes to the left of the vertical line are those explicitly included in the null space.
}
\end{figure}

Intuitively, it is from this property that we can understand how multigrid algorithms can outperform solvers using algorithm deflation. Deflation removes only those modes that are explicitly known, which improves the condition number of the remaining space. Multigrid acceleration includes these explicit modes, but also includes a large number of modes formed by sums over blocks of the low modes, a much larger space, as shown on the left-hand side of Fig.~\ref{fig:nullspace}. These additional modes capture much of the low-lying space just above the explicitly included modes. We show the fraction of the lowest modes of an example Dirac operator that are included in the coarse space on the right-hand side of Fig.~\ref{fig:nullspace}. Each of the 16 lowest modes (marked by the vertical line) are explicitly included at 100\% level, and additionally, at least 90\% of the next 22 modes are included by the blocking at the coarse level.

One drawback of a method relying directly on a determination of the low eigenmodes of the operator is that solving for these modes is often quite expensive. In addition, it is difficult to tell a~priori how many vectors will be needed to remove the critical slowing. We can solve both of these problems by using an adaptive algorithm. In an adaptive algorithm, we apply a Krylov solver to the problem $Ax=0$, where $x$ is initialized to some random vector. After some number of iterations, the vector $x$ will contain primarily the modes which the Krylov solver is not doing a good job of removing (since the true solution is exactly zero). Using this initial vector (or some batch of vectors found in this way), we construct a multigrid algorithm. This multigrid algorithm can itself be applied to the same problem and will fail to remove a different set of vectors. The process may be repeated for as long as necessary until the critical slowing is removed.

Applying a multigrid algorithm to the domain-wall fermion (DWF) Dirac operator poses some unique challenges. Firstly, since the DWF formulation adds a fifth dimension, we must consider how to treat this new dimension, either using the same kind of blocking or giving it some special treatment. Although naively, the fifth dimension appears to be very similar to the usual four dimensions of spacetime, the same physical properties of the DWF operator that make it useful as a nearly chiral fermion formulation also make the fifth dimension special. While the four spacetime dimensions will appear in the structure of the DWF operator's eigenvectors in complicated ways, the low modes will all have the same fifth-dimensional structure: exponentially bound to one side of the domain wall or the other depending on their chirality. Given this simple structre, including the fifth dimension at the coarse level seems unnecessary; all low modes will have the same exponential form in this dimension. Taking advantage of this, we make the coarse blocks cover the entire fifth dimension in all cases. This will greatly accelerate the coarse operators relative to the fine operator while still allowing us to describe every fifth-dimensional behavior relevant to the low modes of the DWF operator.


\section{Application and Testing}\label{sec:num}

In order to demonstrate the success of the multigrid algorithm, we will compare its performance to that of the conjugate-gradient (CG) algorithm applied to the squared DWF Dirac operator. We will apply a quality metric counting the number of times the fine operator is applied. This will somewhat overrate the performance of the multigrid algorithm, since the coarse operators are much faster than the fine operators but not zero cost. However, it is unclear at this time what discount should be associated with an efficient implementation of the coarse operator.

To evaluate the effectiveness of the multigrid as an acceleration technique, we will use a 2-level V-cycle as a preconditioner to a Krylov solver. Since the V-cycle will produce a different approximation to the inverse of the Dirac operator during each evaluation, we should select an outer Krylov solver that can be flexibly preconditioned; generalized conjugate residual (GCR) is well suited to this purpose.

\begin{figure}
\begin{center}
\includegraphics[width=0.45\textwidth]{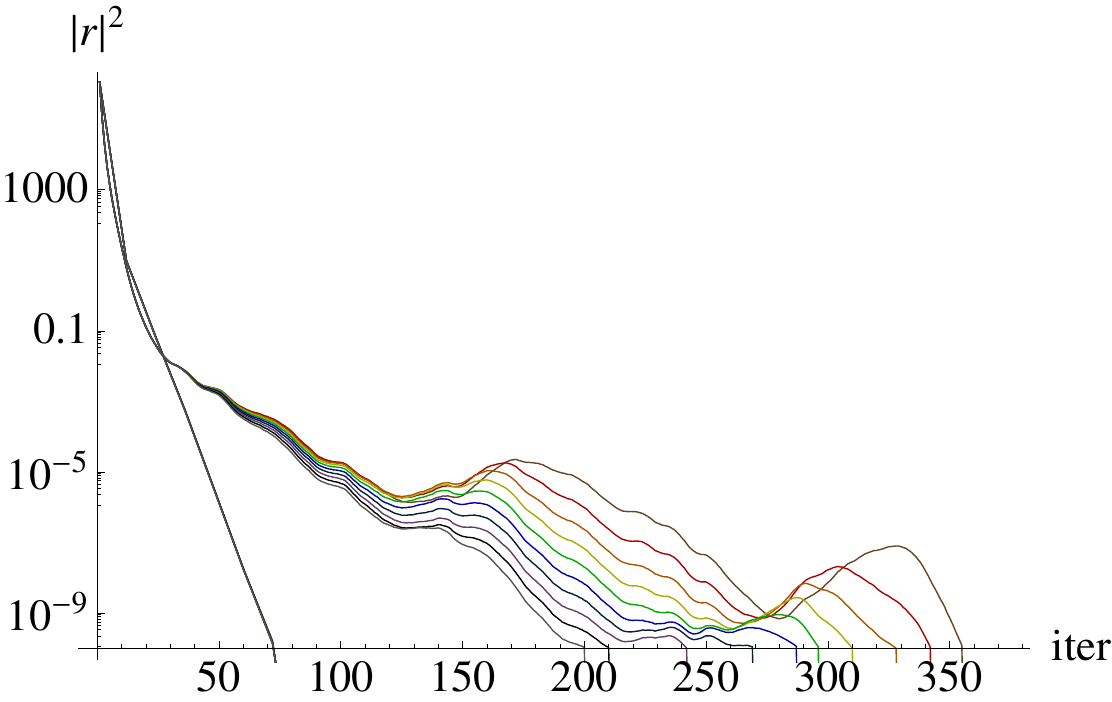}
\includegraphics[width=0.45\textwidth]{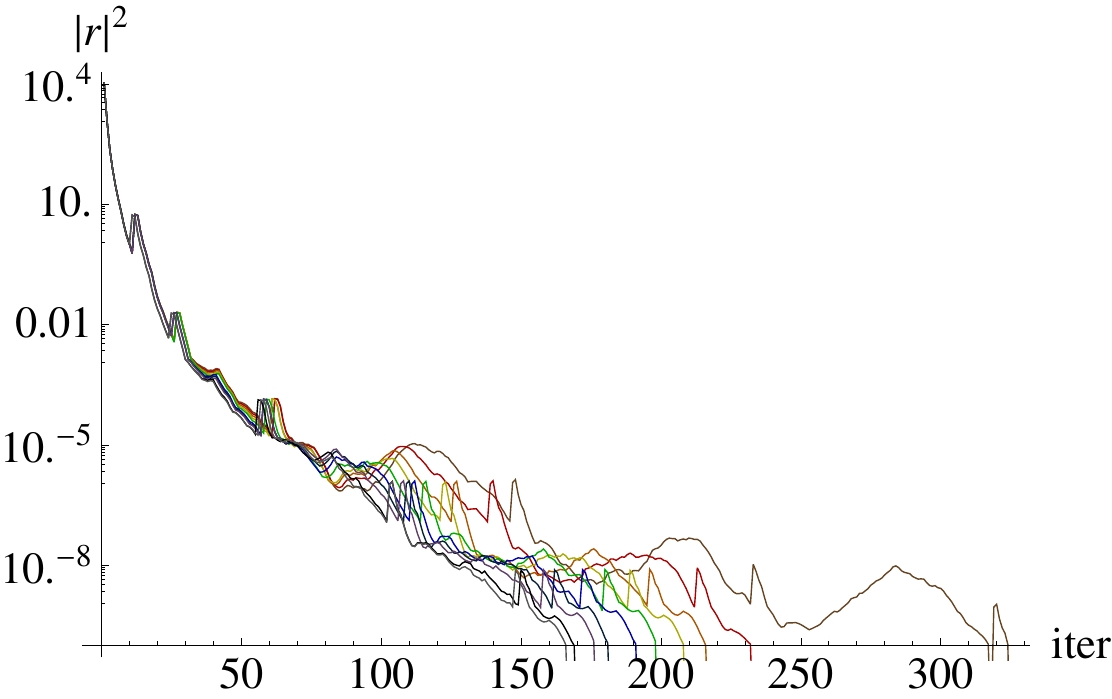}
\end{center}
\caption{\label{fig:2dresult} (Left) Squared residual as a function of the number of applications of the squared fine-lattice DWF Dirac operator on a $20^2\times 8$ U(1) quenched test lattice. The dashed lines are conjugate gradients (CG), while the solid lines are multigrid accelerated. The colors indicate different quark masses, 0.01 to 0.10 in steps of 0.01. (Right) Coarse-level residual as a function of the number of applications of the coarse-lattice operator.
}
\end{figure}

We first apply the two algorithms to a two-dimensional $20^2\times 8$ U(1) quenched lattice, which provides a simple example of the effectiveness of the multigrid algorithm in a context similar to full QCD, since the system exhibits confinement and critical slowing. We use a null space consisting of 16 low vectors and a block size of $4^2\times 8$. On the left-hand side of Fig.~\ref{fig:2dresult}, we show the convergence of CG as dashed lines. The different colors correspond to different quark masses (according to resistor code: brown = 0.01, red = 0.02, etc.); the critical slowing is clearly evident as the quark mass approaches zero. The convergence of the multigrid algorithm, shown as solid lines, demonstrates no such slowing; all the quark masses converge at the same rate and much faster than the CG.

On the right-hand side of Fig.~\ref{fig:2dresult}, we show the convergence of the coarse-level residual in the multigrid algorithm. Unlike the fine-level residual, this does show critical slowing; however, this is not problematic, since the cost of applications of the coarse operator are so small. If the number of iterations needed at this level does become problematic, a third coarser level may be introduced.

\begin{figure}
\begin{center}
\includegraphics[width=0.45\textwidth]{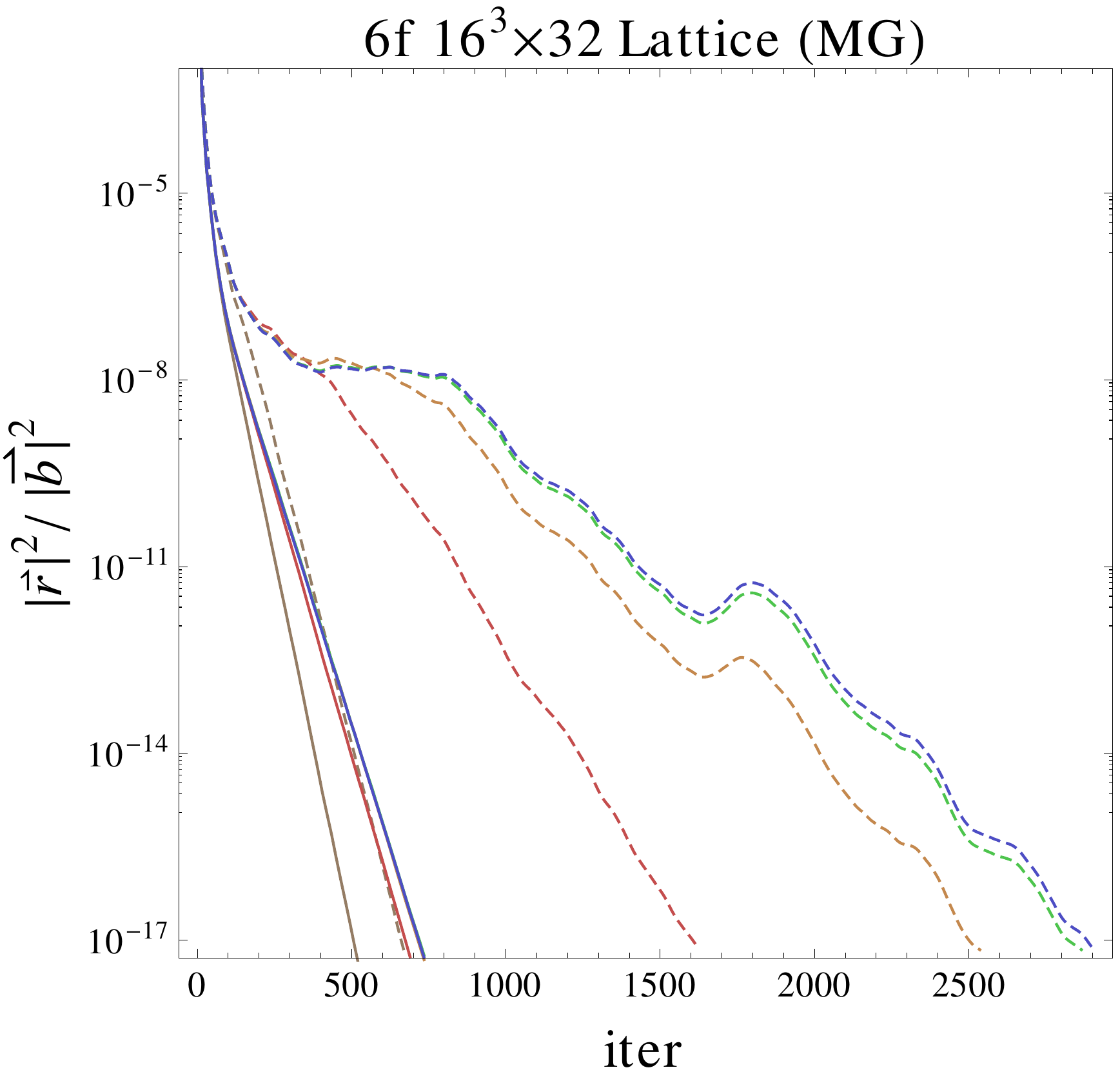}
\includegraphics[width=0.45\textwidth]{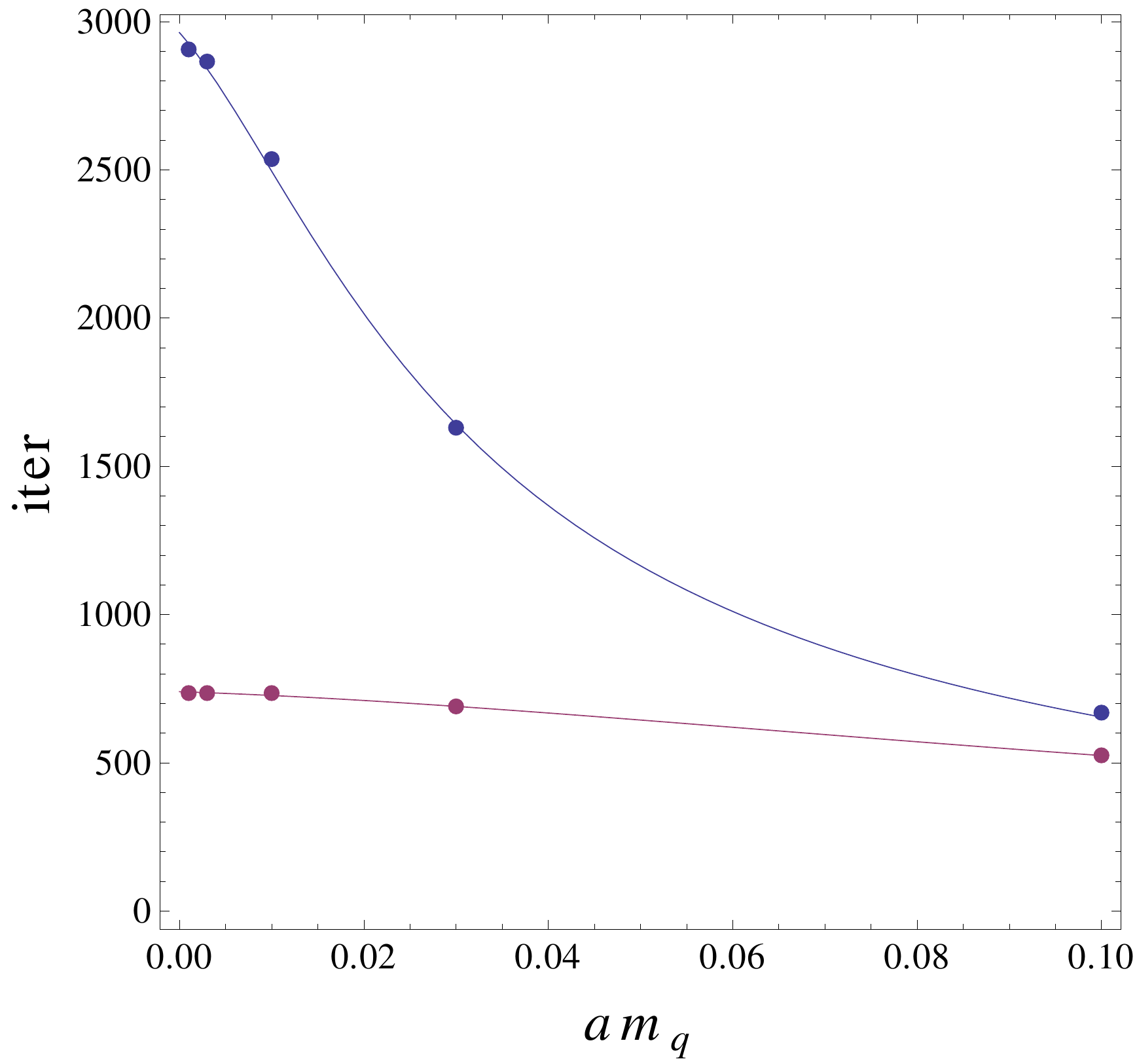}
\end{center}
\caption{\label{fig:6fresult} (Left) Squared residual as a function of the number of applications of the squared fine-lattice DWF Dirac operator on a 6-flavor $16^3\times 32\times 8$ technicolor lattice. The dashed lines are conjugate gradients (CG), while the solid lines are multigrid accelerated. The colors indicate different quark masses: $\{0.1,0.03,0.01,0.003,0.001\}$. (Right) Number of iterations needed to converge for CG (blue) and multigrid (purple) algorithms as a function of quark mass.
}
\end{figure}

Following up this success, we apply the two algorithms to a 6-flavor $16^3\times 32\times 8$ lattice generated by the Lattice Strong Dynamics (LSD) Collaboration\cite{Appelquist:2009ka}. These moderately sized production lattices represent a real-world situation in which the multigrid algorithm would be beneficial. We use 24 null vectors and a blocking of $4^4 \times 16$. As shown on the left-hand side of Fig.~\ref{fig:6fresult}, the multigrid algorithm again nearly eliminates the critical slowing. The amount of speed-up is clear on the right-hand side of Fig.~\ref{fig:6fresult}, about a factor of 4. We expect that the algorithm will perform even better on larger lattices.

\section{Conclusion}\label{sec:end}

In this paper, we have demonstrated how multigrid techniques may be applied to the domain-wall fermion action in lattice gauge theory. The combination of adaptive algebraic multigrid with certain simple strategies specific to the DWF operator have allowed us to remove the critical slowing from this important and expensive problem. As a multigrid algorithm, our suggested strategy is naturally recursive, eliminating the costly volume scaling that would be present in similar acceleration techniques, such as eigenvector deflation. One potential drawback to the method presented here is that it uses a coarsening of the squared operator rather than a coarse version of the operator itself. This means that the coarse operator will have a larger stencil than the fine operator. That is, the value of a particular site following a fine operation depends on the local value and its 10 neighbors at unit distance in the five dimensions, but the value following a coarse operation depends on 17 sites at up to two hops away. This increases the amount of storage required for the coarse operator and means that standard libraries that treat operators having one-hop stencils cannot be applied. We believe that it may be possible to coarsen the unsquared operator directly and are investigating such methods at this time.

\section*{Acknowledgments}
SDC is supported by U.S. Dept. of Energy grants DE-FG02-91ER40676 and DE-FC02-06ER41440, and NSF grant OCI-0749300. 

\bibliographystyle{apsrev}
\bibliography{2011LatCon-proc}

\end{document}